# Optical readout: a tool for studying gas-avalanche processes

A. Rubin[*], L. Arazi, S. Bressler, A. Dery, L. Moleri, M. Pitt, D. Vartsky[†], and A. Breskin

*Department of Particle Physics and Astrophysics*
*Weizmann Institute of Science, 76100 Rehovot, Israel*
E-mail: adam.rubin@weizmann.ac.il

ABSTRACT: Optical recording of avalanche-induced photons is an interesting tool for studying basic physics processes in gaseous detectors. In this work we demonstrate the potential of optical readout in avalanche-propagation investigations in Thick Gas Electron Multipliers (THGEMs) operated with Ne/CF$_4$ (95/5). We present the results of direct measurements, with single- and cascaded-THGEM detectors irradiated with soft x-rays, of the hole-multiplicity and avalanche asymmetry within holes, as a function of detector parameters. Further study directions are discussed.

KEYWORDS: Charge transport and multiplication in gas; Gaseous imaging and tracking detectors; Micropattern gaseous detectors (MSGC, GEM, THGEM, RETHGEM, MHSP, MICROPIC, MICROMEGAS, InGrid, etc); Scintillators, scintillation and light emission processes (solid, gas and liquid scintillators).

---

[*] Corresponding author.
[†] On leave from Soreq NRC, Yavneh, Israel.

# Contents

## 1 Introduction

Optical readout is the method of recording photons emitted in electron transport and multiplication processes in gaseous detectors. It offers some advantages over electronic charge readout: it allows the recording of a large field of view with high spatial resolution without using complex data acquisition and anode geometry; it is unaffected by electromagnetic interference that adds background noise to charged readout; and it can be implemented with simple, off-the-shelf CCD/CMOS cameras and image intensifiers. For a review on optical imaging detectors see [1] and the references therein.

Typically, optical readout schemes utilize electron multipliers (e.g. Gaseous Electron Multiplier, GEM [2]) to convert deposited charge in the system into photons which can be recorded. Optical readout has been used for various applications such as single-photon localization in Cherenkov Ring Imaging [3], thermal neutron imaging [4–6], and rare-event tracks in Time Projection Chambers [7,8]. Alternatively, it has been used to efficiently identify defects in large area GEMs and Micro-Strip Gas Chambers (MSGC) [5]. Lately, it has been proposed to optically record avalanches from Thick Gas Electron Multiplier (THGEM [9]) holes in liquid argon [10] and in the vapor phase of liquid xenon [11], with G-APD sensors. In a recent work [12] it has been proposed to optically record avalanche- and electroluminescence-photons from cascaded hole-multipliers in the noble liquid itself, with gaseous photomultipliers [13].

Photons are created in gaseous detectors by excitation and de-excitation of the gas atoms and molecules [14]; in particular during the avalanche process [15]. The emission yield (photons/avalanche electron) and the emission spectrum are a function of the gas, its pressure and the electric field; the latter is dictated by the detector geometry. Examples of gases with a high photon yield are mixtures of Ar with triethylamine (TEA) [16], some gases with Tetrakis dimethylamino ethylene (TMAE) [3], Ar or Ne with $CF_4$ [17,18], and $N_2$ with $CO_2$ [19]. $CF_4$, used in this work, has some very convenient properties: it is a suitable quencher, absorbing secondary vacuum ultra violet (VUV) photons while copiously emitting photons in the visible band—making the use of standard optics possible. For more information on the scintillation properties of $CF_4$ and its uses in gas detectors see [17,18,20,21].

Besides the aforementioned applications, optical readout could be useful for studying the physics of basic electron transport and multiplication processes. It could provide useful information on the avalanche shape, size, diffusion, and secondary effects. In this work, we investigated the potential of the optical readout method for studying basic avalanche processes in THGEM detectors.

Two independent studies were performed. First, we used a non-collimated x-ray source to irradiate THGEM detectors—with different numbers of cascaded elements—and measured the hole-multiplicity (number of activated holes per photon interaction). Second, we used a collimated x-ray source and measured the asymmetric development of electron avalanches within a hole, as function of the irradiation geometry. Notably, while we demonstrated the validity of the method in THGEM detectors, investigated by us for a variety of applications [9], it should be applicable to a large range of other electron multipliers.

## 2 Experimental setup and methodology

The experimental setup can be divided into three parts: the radiation source, the irradiated detector emitting the avalanche photons, and the optical readout system recording these photons. Figure 1 shows the detector scheme (here with a double-THGEM assembly) and the optical chain. The radiation sources used in both experiments were significantly different; therefore we defer their description to section 2.3.

### 2.1 Detector setup

In this work various THGEM configurations were used. The THGEM is an electron multiplier in which avalanche multiplication develops within sub-millimeter diameter holes, mechanically drilled in a standard two-sided copper-clad printed circuit board (PCB). Radiation-induced ionization electrons are focused into the holes and multiplied in an avalanche process under the high electric field set by the potential difference between the THGEM faces. Very large gains, exceeding $10^6$, can be reached by cascading a few THGEM elements [22]. The reader is referred to [9] for a review on THGEM principles and applications.

The detector (single-, double- or triple-THGEM) was assembled in an aluminum chamber; it was continuously flushed with 1 atm of Ne/$CF_4$ (95/5) at typical flow rates of a few tens of sccm, using an MKS 146 mass-flow control system. The chamber had a 50 μm thick Kapton x-ray window, and on the opposite end—a 43 mm diameter, 3.5 mm thick, quartz window through which the emitted photons were collected by the optical chain. Yellow LEDs (with a guard ring of inner diameter 29 mm) were placed outside the window for alignment purposes. The parameters of the $30 \times 30$ mm$^2$ THGEM electrodes employed in this work are summarized in table 1.

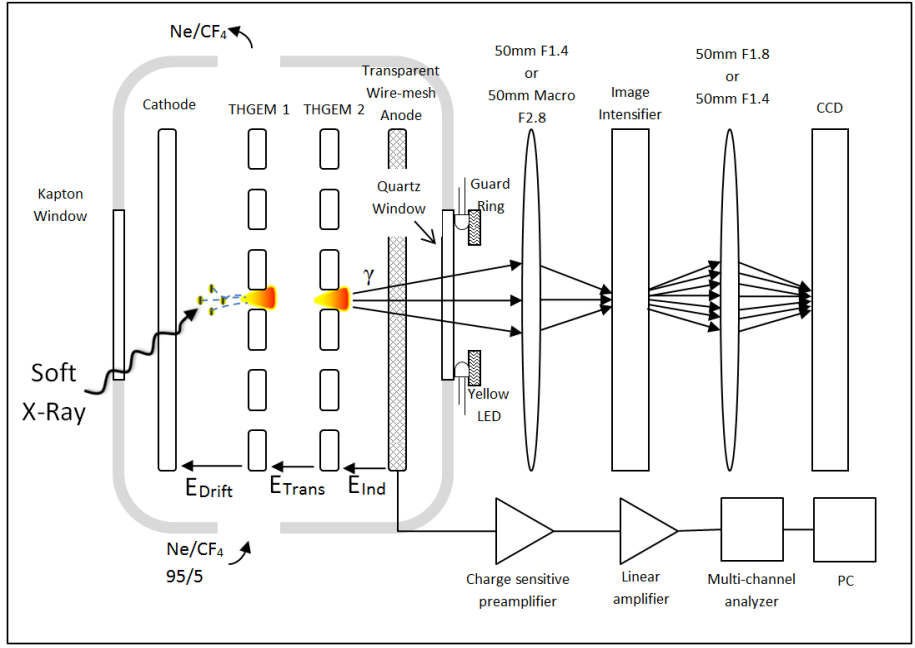

Figure 1: Experimental setup. The detector shown combines a cathode, a double-THGEM, and a wire-mesh anode. The light emitted by radiation-induced avalanches is recorded through a quartz window, by an image intensifier viewed by a lens, amplified and then focused by a second lens onto a CCD camera.

Table 1: THGEM geometries investigated.

| THGEM # | Pitch (a) [mm] | Hole diameter (d) [mm] | Thickness (t) [mm] | Rim size (h) [mm] |
|---|---|---|---|---|
| 1 | 1 | 0.5 | 0.4 | 0.1 |
| 2 | 1.3 | 0.7 | 0.4 | 0.1 |
| 3 | 1.5 | 0.8 | 0.8 | 0.1 |
| 4 | 1.3 | 0.7 | 0.8 | 0.1 |

A cathode electrode (aluminized Mylar or a THGEM with its faces interconnected) placed in front of the drift gap (figure 1) provided a drift field to drive the electrons into the multiplier; a wire-mesh (of 82% optical transparency) following the induction gap collected the avalanche charge while allowing the emitted photons to pass through. The electrodes were biased with CAEN N471A power supplies through low-pass filters. The avalanche charge was collected and read out through an Ortec 142 charge sensitive preamplifier. The signal was then amplified with an Ortec 570 linear amplifier and the spectrum and rates were measured with an Amptek 8000A multi-channel analyzer (MCA).

## 2.2 Optical setup

Avalanche photons were collected and imaged by the intensified CCD camera system shown schematically in figure 1. Photons traversing the chamber's quartz window were collected and focused with a lens (Nikon AF Nikkor 50 mm 1:1.4D or Sigma 50 mm F2.8 EX DG Macro) onto a Proxitronic image intensifier, powered by a Topward 6303D dual-tracking DC power supply.

The intensifier, with a 25 mm diameter S20-photocathode, was type 2563MZ-V 100N dual micro channel plate (MCP), in a V-Stack assembly. The intensifier's P43 phosphor screen had a 1 ms decay-time. The image intensifier was operated at 1640 V (~90% of the maximal allowed amplification voltage).

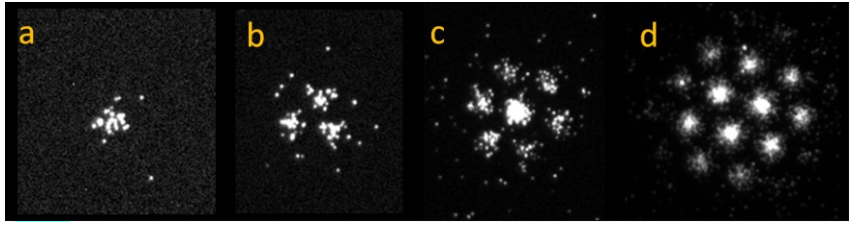

Figure 2: Examples of non-collimated 5.9 keV x-ray induced single-event avalanches recorded in different detector configurations of the setup shown in figure 1. a) Single-THGEM with a reversed drift field of 0.3 kV/cm and gain ~$10^4$; b) Single-THGEM ($E_{drift}$ 0.5 kV/cm, gain ~$10^4$); c) Double-THGEM with 8 mm transfer gap ($E_{drift}$ 0.5 kV/cm, $E_{trans}$ 0.5 kV/cm, gain ~$5 \times 10^5$); d) Triple-THGEM with 8 mm and 10 mm transfer gaps ($E_{drift}$ 0.5 kV/cm, $E_{trans}$ 0.5 kV/cm, gain ~$10^7$). The images are unprocessed, but the contrast has been adjusted to improve visibility. THGEM type 1 (table 1); gas: Ne/$CF_4$ (95/5).

The intensifier's phosphor screen was then imaged onto a Finger Lakes Instrumentation CCD camera model MX0013307 with a 50 mm lens (Canon FD 50 mm 1:1.8 or Nikon AF Nikkor 50 mm 1:1.4D); CCD signals were then recorded. The images were acquired with a 50 ms exposure time and were taken at a rate of ~3-5 Hz. The detected x-ray rates of 4-30 Hz were set to reduce recorded-events overlap in a given frame to values < 10%. Due to the short exposure times the CCD's thermal noise was insignificant, therefore cooling was not necessary. In order to determine the geometrical position of the detector's holes, it was illuminated from outside the chamber with yellow LEDs (figure 1). Examples of images obtained in different configurations can be seen in figure 2.

The photon-yield reaching the CCD camera was estimated based on the following data: avalanche photons were transmitted through the anode mesh, of 82% optical transparency; they were collected by the first lens (imaging the detector onto the image intensifier) at a solid angle of ~7%, with the lens transmission of ~90%; the S20 photocathode had ~6.5% quantum efficiency at the main Ne line in the Ne/$CF_4$ spectrum (586 nm, [23]); the double MCP had a gain of ~$10^4$; the phosphor screen released ~185 photons/e at a wavelength of ~545 nm; it was viewed by the second lens (imaging the phosphor screen onto the CCD) with a solid angle of ~1.5% with a transmission of ~90%; the CCD had a quantum efficiency of ~65% at 545 nm and 2.5 electrons per analog/digital unit (ADU).

In our measurements, 8 keV x-rays at a detector gain of ~$10^4$ yielded $7.8 \pm 4 \times 10^5$ ADU (mean ± standard deviation). This indicates that the secondary photon yield of Ne/$CF_4$ (95/5) in THGEM at gain $10^4$ is $1.6 \pm 0.8 \times 10^{-2}$ ph/e$^-$. This is ~25% of the 0.06 ph/e$^-$ estimated by Tokanai et al. [23], though for a different Ne/$CF_4$ (90/10) mixture at a gain of ~$10^4$ in a capillary plate (CP) detector. The use here of the manufacturer's nominal MCP-gain value and the different field geometry in a THGEM and a CP, could also explain the four-fold difference in the light yield.

The images recorded by the CCD contained two types of noise: CCD thermal pixel-noise and the image intensifier's noise. Figure 2 shows that avalanches, being formed of densely distributed spots, are clearly distinguishable whereas the random noise is sparse. The image processing was done using built-in functions of the Mathworks Matlab® R2012a Image Processing Toolbox [24]. The CCD noise was eliminated by subtracting an image taken with the shutter closed, setting a threshold on the intensity at the mean plus twice the standard deviation (leaving a binary black and white image), and applying a $3 \times 3$ pixel median filter.

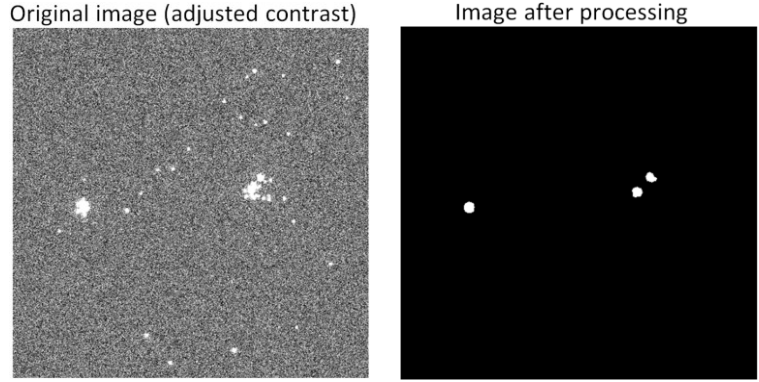

Figure 3: Example of the image before (left, containing CCD thermal noise and single-spot image intensifier noise) and after (right) processing. Three holes were identified: one event with a hole-multiplicity of two and the other with a multiplicity of one.

The image intensifier noise was eliminated by applying a morphological closing filter—using a disc-shaped structuring element with a radius set to be roughly half the radius of a hole—then discarding all information outside the holes and applying an area criterion: if the multiplicity was >1, a spot was kept only if its area was larger than 50% of the area of a hole; if the multiplicity was 1, a spot was kept only if its area was larger than 80% of the area of a hole. Only events that were far from the edges of the frame were considered, to avoid underestimating the multiplicity. Figure 3 shows an example of an image before and after processing.

## 2.3 Study description

The two studies performed were different conceptually. In the first, an un-collimated x-ray source was used to irradiate a detector comprising one, two, or three cascaded THGEMs to provide varying hole-multiplicities. In the second, a collimated source was scanned across the hole of different THGEM types and the avalanche asymmetry within the hole was studied.

### 2.3.1 Hole multiplicity for different detector configurations

The detector was irradiated with 5.9 keV x-rays from an un-collimated $^{55}$Fe source. The detector consisted of one or more THGEMs separated by transfer gaps, and a metallic mesh to bias the induction gap and collect the charge. The THGEMs used to multiply the electrons were all of type 1 (table 1). The configurations and fields are given in table 2. Several thousand CCD frames were captured and analyzed in each configuration.

Table 2: The configurations used to study the hole-multiplicity with a THGEM of type 1 (table 1).

| Configuration | Number of cascaded THGEMs | THGEM voltages [V] | Gain | Drift gap [mm] | Drift field [kV/cm] | Transfer gaps [mm] | Transfer fields [kV/cm] | Induction gap [mm] | Induction field [kV/cm] |
|---|---|---|---|---|---|---|---|---|---|
| a | 1 | 830 | ~$10^4$ | 10 | Inverted 0.3 | ----- | ----- | 5 | 0.5 |
| b | 1 | 830 | ~$10^4$ | 10 | 0.5 | ----- | ----- | 5 | 0.5 |
| c | 2 | 650,750 | ~$5 \times 10^5$ | 10 | 0.5 | 8 | 0.5 | 5 | 0.5 |
| d | 3 | 700,650,600 | ~$10^7$ | 10 | 0.5 | 10,8 | 0.5, 0.5 | 5 | 0.5 |
| e | 3 | 450,600,700 | ~$10^6$ | 10 | 0.5 | 2,2 | 0.5, 0.5 | 5 | 0.5 |

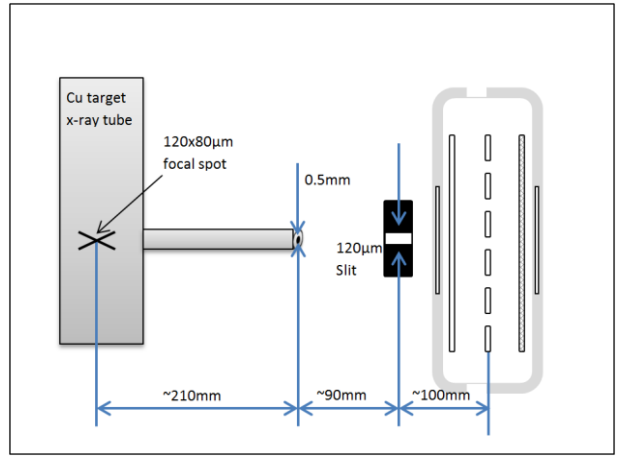

Figure 4: Experimental setup for scanning across a THGEM hole with a collimated x-ray beam (dimensions not to scale).

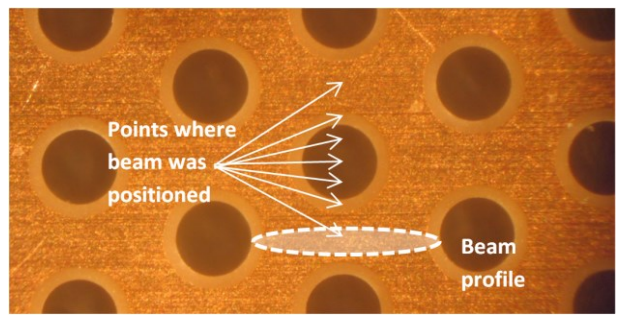

Figure 5: Irradiation locations throughout x-ray beam scanning across a THGEM-hole.

### 2.3.2 Scan across a hole with a collimated x-ray beam

The detector was irradiated with a collimated 8 keV x-ray beam (figure 4). The detector was precisely displaced allowing for beam positioning at different locations relative to a hole center (in ~140 μm steps) as is illustrated in figure 5.

The x-ray beam was prepared by using an Oxford Instruments copper target x-ray tube, with a focal spot size of $80 \times 120$ μm$^2$ and filtered with ~30 μm thick copper foils; the beam was collimated with a 0.5 mm diameter circular aperture placed ~210 mm from the focal point and a 120 μm slit placed ~90 mm from the circular collimator. The detector was placed ~100 mm from the slit (figure 4).

The beam profile, of 330 μm by 730 μm (FWHM), was measured by displacing an Amptek XR-100CR silicon x-ray detector (attached to a 200 μm slit) relative to the collimated x-ray beam[1]. The beam position relative to the THGEM detector was determined by placing a ZnS(Ag) screen in the focal plane of the optical system (instead of the chamber) and fitting the x-ray induced fluorescent spot to a Gaussian. Only the THGEM chamber was displaced in the experiments, therefore the pixel representing the position of the beam remained fixed.

---

[1] The stated beam size is corrected for the scanning slit's width.

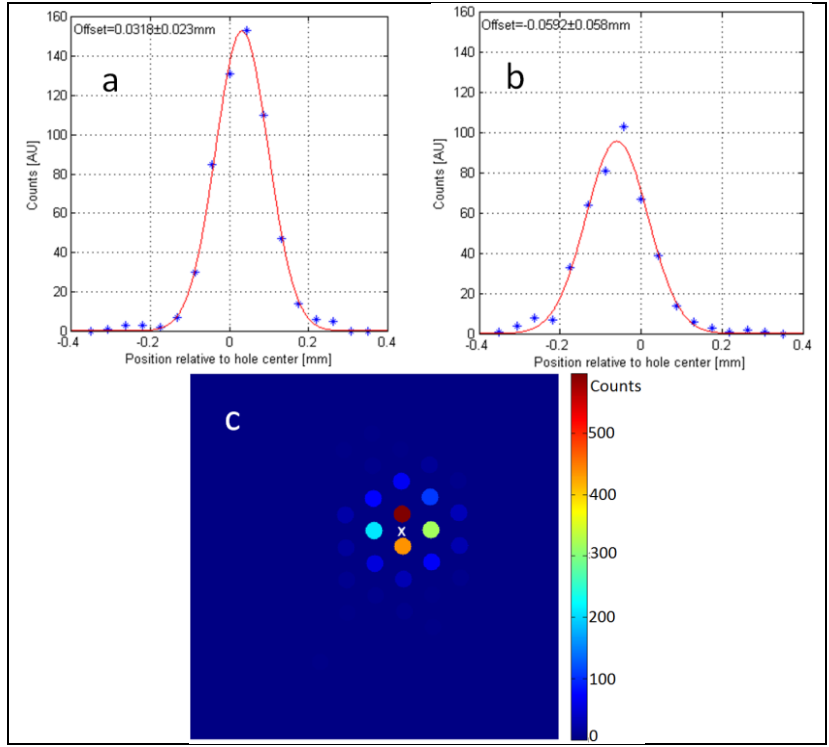

Figure 6: Scanning across a THGEM hole (see figure 5). Shown are the analysis results with the x-ray beam positioned in between two holes. a) and b): histograms of the centroid of the light distribution in the vertical direction in holes located above and below the beam respectively. The distribution in both holes is offset in the direction of the beam. c) Color coded image, with the color corresponding to the number of times a hole participated in an event (the cross being the predicted beam position deduced from the image on the phosphor screen). Detector gain $10^4$.

For each image, the center of gravity of the light emitted from the scanned hole, after subtracting the mean intensity, was calculated. The centers of gravity were aggregated into a histogram, one for the x coordinate and one for the y coordinate, and fit to a Gaussian; its mean being the offset of the centroid position from the center of the hole. An example where the beam was positioned in-between holes is given in figure 6a and 6b; in both holes the light emission was asymmetric, shifted towards the beam's position. Figure 6c is a histogram showing the number of times each hole participated in an event. As expected, the holes closest to the estimated beam position are most active, with the activity decaying with distance. This was performed as a consistency check for the image processing algorithm.

The study was conducted on a detector with a 0.5 kV/cm drift field applied over a 5 mm gap, a single-THGEM multiplier and an anode mesh; a 0.5 kV/cm induction field was set over the 5 mm induction gap. Several THGEM electrodes were investigated, with different parameters as listed in table 1.

A second variant of the avalanche-asymmetry study was performed by using a THGEM of type 1 (table 1), biased for an effective gain of ~$10^4$, keeping the induction field at 0.5 kV/cm and varying the drift field from 0.2 to 1.5 kV/cm. The purpose was to investigate the effect of the drift field on the degree of asymmetric avalanche formation. The configurations tested are given in table 3.

Table 3: The drift field configurations of a single-THGEM (type 1, table 1) detector scanned across with a collimated x-ray beam (figure 4).

| Configuration | Drift gap [mm] | Drift field [kV/cm] | Induction gap [mm] | Induction field [kV/cm] |
|---|---|---|---|---|
| f | 5 | 0.2 | 5 | 0.5 |
| g | 5 | 0.5 | 5 | 0.5 |
| h | 5 | 1.0 | 5 | 0.5 |
| i | 5 | 1.5 | 5 | 0.5 |

### 2.3.3 High statistics runs and simulations

In order to estimate the distribution of events when irradiating at different locations relative to the hole, two high-statistics runs were conducted with a THGEM of type 1 (table 1) in configuration g (table 3), collecting 15,000 events with the beam positioned at the center of a hole, and 20,000 events with the beam positioned in between adjacent holes. This was compared to a simulation of the same THGEM and field configuration, with 800 V across the THGEM, using ANSYS [25] for the field calculations and Garfield [26] for electron transport. In the simulation, 8 keV electrons (drawn from a uniform distribution along z and a two-dimensional Gaussian distribution with FWHM of 730μm and 330μm in x,y) were generated for the two beam positions, with isotropic initial directions. The electrons created—following gas ionization events by the primary 8 keV electron—were allowed to drift through the THGEM holes, with their hole-crossing position (x,y) recorded. No electron multiplication was allowed. For each beam position $5 \times 10^4$ 8 keV electrons were released. The center of gravity of each event was calculated as the center of gravity of the x,y coordinates of the electrons exiting the THGEM (from several holes).

## 3 Results

### 3.1 Hole multiplicity

Figure 7 shows the measured multiplicity distribution for the configurations given in table 2. The figure indicates that events from the un-collimated 5.9 keV x-rays were found to span 1-20 holes depending on the detector configuration. The inverted drift field (resulting in conversion close-to and within holes) resulted in a multiplicity of 1 in 86% of the events. While the gain was not kept fixed between the different configurations, one can see that qualitatively the multiplicity is strongly affected by the number of stages and by the size of the transfer gaps; reducing the transfer gap in configuration e (table 2) reduced the multiplicity considerably.

### 3.2 Scan across a hole

Figure 8 shows the results of measurements of the avalanche displacement as function of the beam position, for the electrode geometries provided in table 1. Displacing the beam relative to the hole center shifted the light-emission distribution from the hole (figure 8a). As expected, no shift was apparent in the direction perpendicular to the scan direction (figure 8b). Changing the hole-geometry and electrode thickness had a minor effect on the displacement of the avalanche (figure 8). Figure 9 depicts the avalanche displacement as a function of the drift-field measured with a detector (type 1, table 1) in the configurations described in table 3. It shows that changing the drift-field at constant gain has small influence on the distribution of light emission within the hole, but that the shift is stronger for higher drift-fields. A seven-fold increase in the drift-field (at the same gain) caused a maximal shift of ~0.1 mm in both cases.

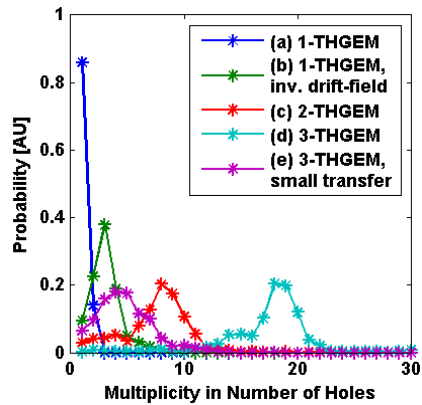

Figure 7: Multiplicity distributions for different configurations (letters in the legend refer to table 2) of the THGEM detector (type 1, table 1). Gas: Ne/CH$_4$ (95/5), $E_{drift}=E_{trans}=E_{ind}=0.5$ kV/cm, Source: 5.9 keV x-rays.

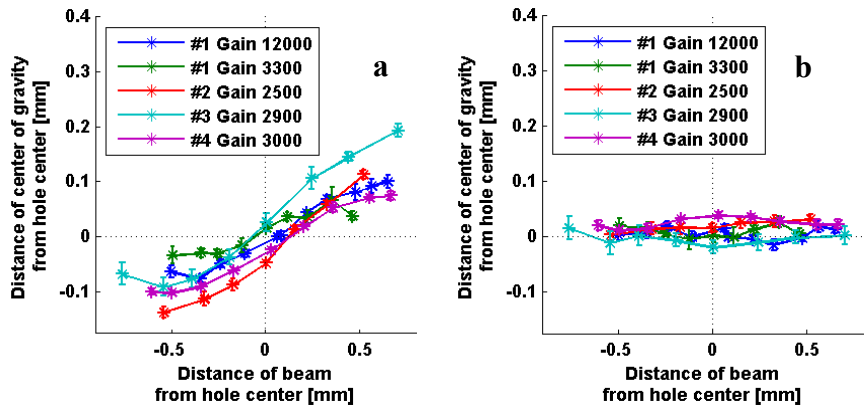

Figure 8: Measurements results of the displacement of the avalanche center of gravity vs. the irradiation position, obtained by scanning different THGEM electrodes (parameters in table 1 and gains indicated in the figure) with a collimated 8keV x-ray beam across a single hole. a) In the scan direction (figure 5). b) Perpendicular to the scan direction. Error bars are the 95% confidence level of the Gaussian fit. The gains and THGEM geometries are provided in the figure; gas Ne/CF$_4$ (95/5).

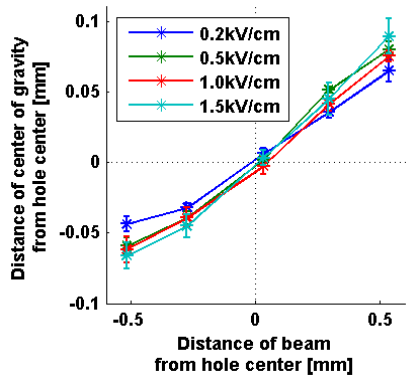

Figure 9: Measurements results of the avalanche displacement vs. irradiation position, obtained by scanning a THGEM (type 1, table 1) electrode with a collimated 8 keV x-ray beam across a single hole. Here the drift field was varied (indicated in the figure). The detector was operated at a gain of $10^4$ in field configurations according to table 3. Error bars are the 95% confidence level of the Gaussian fit.

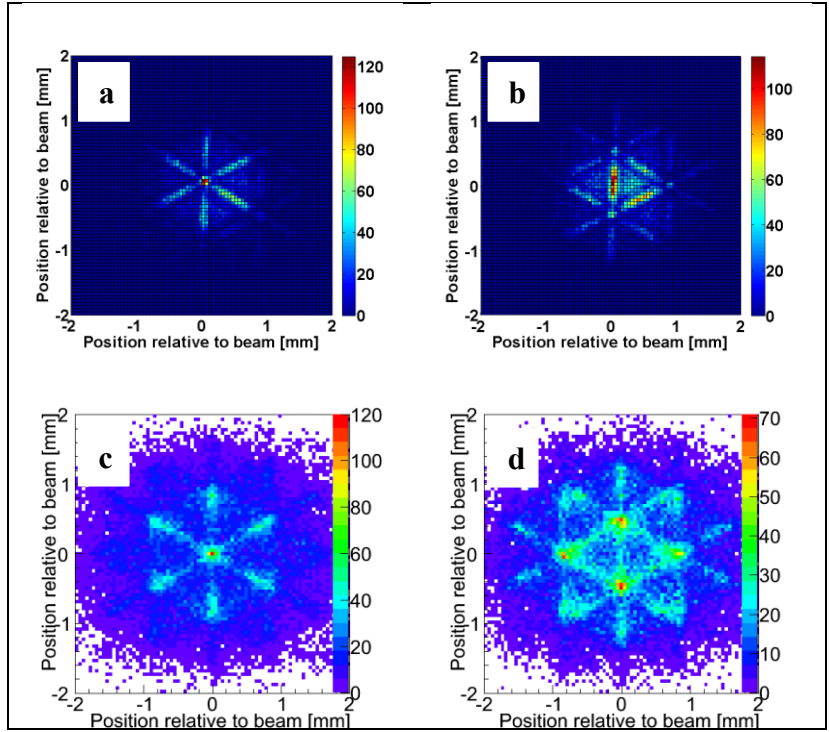

Figure 10: Reconstruction of the points of interaction, from the center-of-gravity of the light emitted from several holes, when irradiating a THGEM type 1, configuration g (table 3) at the center of a hole and in-between holes. a) and b) are experimental results (with 8 keV x-rays) at gain $10^4$; c) and d) are simulation results (with 8 keV electrons). See text for explanations of the figures, and section 2.3.3 for explanations of the experiment and simulation.

### 3.3 High statistics runs and simulations

Figure 10 depicts the results of the high statistics runs (>$10^4$ events of 8 keV x-rays), taken at a gain of $10^4$ at the center of a hole and in-between holes, as described in section 2.3.3. It shows how the center-of-gravity of the event (experiment and simulation) is distributed—for irradiation at the center of the hole and in-between holes, respectively. Figure 10a shows the distribution of the experimental center-of-gravity for irradiation at the center of the hole. Most events' center-of-gravity fell at the center of the hole. However many events divided between two adjacent holes—causing the center-of-gravity to fall in between them—resulting in the "star shapes" visible in the figure. Figure 10b shows the results of the detector irradiation in-between holes; the center-of-gravity fell primarily in-between them (the intense line connecting the holes). The results of the simulation of irradiation with a Gaussian beam (see section 2.3.3) are presented in figure 10c and 10d. For irradiation at the center of a hole, the simulation (figure 10c) shares several key features with the experiment (figure 10a); namely the reconstructed center-of-gravity lies along the lines connecting neighboring holes and there is almost no reconstruction just outside the central hole. However, for irradiation in-between holes, the simulation (figure 10d) lacks several key features of the experiment (figure 10b); namely the reconstruction is primarily in the holes, while in the experiment the reconstruction was primarily in-between the holes. This may be due to oversimplification in the simulation (disregarding multiplication, energy-inhomogeneity of the incident radiation, and beam misalignment). More detailed modeling was beyond the scope of this work.

Table 4: Radii containing 33, 66 and 95 percent of the center-of-gravity of the events (experiment and simulation) when irradiating a THGEM of type 1 (table 1) in configuration g (table 3) in the center of a hole and in-between holes at a gain of $10^4$. Images are given in figure 10.

|  | Irradiation position (center of hole / in between holes) | Radius containing 33% of events [mm] | Radius containing 66% of events [mm] | Radius containing 95% of events [mm] |
|---|---|---|---|---|
| **Experiment** | Center | 0.45 | 0.67 | 1.4 |
|  | In between | 0.43 | 0.73 | 1.4 |
| **Simulation** | Center | 0.8 | 1.2 | 1.9 |
|  | In between | 0.79 | 1.2 | 1.88 |

In order to quantitatively estimate the precision of event-position reconstruction, table 4 compares the center-of-gravity distributions resulting from the experiments and the simulations. Experimentally 95% percent of the 8 keV events were found to be within a radius of 1.4 mm from the point of irradiation, both when the beam was at the center of a hole and when it was in-between holes. Simulations showed that 95% percent of the 8 keV events were found to be within ~1.9 mm of the point of irradiation. This may be due to the experimental threshold (criteria for identifying a hole that participated in the event), which may cause the long tails of the distributions to be ignored. While there appears to be no difference in table 4 between irradiation at the center and in between holes, the discrepancy in the experiment between the radius containing 66% of the events is not clear, and may be due to misalignment of the beam.

## 4    Discussion

Optical avalanche-recording studies of the number of holes involved in charge multiplication of an x-ray induced event were carried out in different conditions. They indicated that the parameters affecting hole-multiplicity are the size of the initial photoelectron track, the point of interaction, the length of the drift-gap and transfer gap(s) and the values of their corresponding electric fields, as well as the electron-diffusion properties of the gas mixture. For example, $^{55}$Fe-induced 5.9 keV photoelectrons yield on the average 164 electron-ion pairs in Ne/CF$_4$ (95/5) along a track length[2] of ~900 μm. The initial electron cloud diffuses and subsequently reaches several holes. For a triple-THGEM structure with 8-10 mm drift and transfer gaps (configuration d, table 2) a typical event had a FWHM of 2.6 mm. In Ne/CF$_4$ (95/5) the diffusion coefficient at 0.5 kV/cm is ~250 μm/$\sqrt{cm}$, indicating a FWHM of ~1 mm for 3 cm drift (drift gap + two transfer gaps). Clearly this is not the dominant process; however two other effects can influence the event's spread: the discretization of the holes (effectively binning the electrons into bins 1 mm apart); and misalignment of different stages (causing the electrons to be re-binned at each stage). In simulation with Garfield of 5.9 keV electrons in a triple-THGEM with effective gain ~30, drift and transfer fields of 0.5 kV/cm, and drift and transfer gaps of 1 cm, we found the electrons exiting the final stage to have a FWHM of only 1.6 mm. A full study—including exploration of possible misalignment of the stages—is necessary.

---

[2] This was simulated with HEED [27] by releasing 5.9 keV electrons in Ne/CF$_4$ (95/5) and measuring the range of the track created. The range obtained from the approximation $R_p = 0.71 E^{1.72}$ (E in MeV, R$_p$ in g cm$^{-2}$) [28] for pure neon is 1.24 mm.

The inverted drift field in front of a single-THGEM prevented primary electrons created in the drift gap from reaching the multiplication stage; here, only x-ray photons that interacted within the THGEM holes were multiplied, resulting in single-hole illumination, in agreement with our expectation (figure 7). Generally, increasing the number of stages caused higher multiplicity due to diffusion and the discretization of the THGEM holes; however the hole-alignment was not well controlled in the experiment, calling for more detailed studies. A somewhat surprising result was that the triple-THGEM with small transfer gaps had an average multiplicity (~4-5) that was close to that of the single-stage detector (~3). The fact that multi-stage detectors can maintain a relatively low hole multiplicity is encouraging, for applications which call for a small signal spread on the readout plane; with lower voltage per element, cascaded detectors are known for their higher stability during high-gain operation [22].

The results in figure 8 indicate a clear correlation between the position of irradiation and the distribution of light emitted within a hole—similar to avalanche asymmetry in other gas-avalanche detectors, e.g. wire chambers. In all of the THGEM electrodes investigated, the center of gravity of the light within the hole was only mildly displaced from the center (<50% of the radius of a hole). The electric field is known to reach higher values at the edge of the hole [29], indicating higher local multiplication. However, the computation of the electric field lines [29] (drifting the electrons) indicates that electrons do not reach the hole's edge. We conclude from our results (figure 8) that although the electric field is higher near the edges of the hole, the displacement of the avalanche from the hole's center is not a major effect. We draw the same conclusion from the drift-field variations (figure 9): although the drift-field influences the displacement of the avalanche, the effect is rather weak. Our conclusion is that in THGEM detectors, the avalanche develops primarily away from the edge of the hole.

The two-dimensional center-of-gravity distributions of the reconstructed points of radiation interaction display important features of the detector. For example, in figure 10a, the lines connecting adjacent holes indicate that most events result in avalanches shared by two holes. These features require deeper study, as they may depend on the type of algorithms used to identify avalanches; specifically, cutting low-charge events may have an impact on multiplicity estimation. Our measurements agree qualitatively with simulation for irradiation at the center of a hole (figure 10a and 10c), but less so for irradiation in-between adjacent holes (figure 10b and 10d)—the simulation missed the accentuated line connecting the holes. The simulation also showed a broader distribution: 95% of events were contained in 1.9 mm opposed to 1.4 mm in the experiment. This is likely due to the existence of a detection threshold (criteria for identifying a hole that participated in the event) in the experiment which inevitably cuts the tail. Exploring the point-spread-function was beyond the scope of this work; however it is notable that both simulation and experiment were generally consistent with each other, and also with THGEM imaging investigation with 8 keV photoelectrons in neon mixtures [30].

## 5  Conclusions

In this work we demonstrated the use of optical readout to study avalanche processes in THGEM detectors. We showed examples of event hole-multiplicity in various THGEM structures. These results should however be regarded as qualitative, requiring a broader systematic study to evaluate the dependence on the gain, the gap size, the division of gain between stages etc. Optical readout would be an appropriate tool to conduct such a study. Our second demonstration explored the avalanche formation within the THGEM hole, showing that it developed away from the hole-edge. The asymmetrical development of the avalanche could, in principle, be used to obtain improved positioning resolution.

While these demonstrations were made on THGEM electrodes, they are naturally applicable to other avalanche detectors. Among other topics to which optical readout may be applied are: photon-feedback—which would manifest as multiple satellite events in the same frame; defect identification (as was done in [5]); and the study of discharges.

The optical method's advantage over electronic readout is its simplicity; high resolution is achieved with off-the-shelf equipment, simple data acquisition, and full decoupling between the detector and its readout system.

**Acknowledgements**

This work was supported in part by the Israel-USA Binational Science Foundation (Grant 2008246). A. Breskin is the W.P. Reuther Professor of Research in the Peaceful use of Atomic Energy.